\documentclass[journal,twoside]{appolb}

\usepackage{amsmath}
\usepackage{amssymb}
\usepackage{amsthm}
\usepackage{cancel}
\usepackage{cite}
\usepackage{epsf}
\usepackage{subfigure}
\usepackage{epsfig,graphics}
\usepackage{multicol}
\DeclareMathOperator{\prob}{\mathbb{P}}
\DeclareMathOperator{\ex}{\mathbb{E}}

\theoremstyle{plain}

\theoremstyle{definition}

\theoremstyle{remark}
\newtheorem*{remark*}{Remark}

\numberwithin{remark}{theorem}

\def\vec#1{{\bf #1}}

\def\sgn{\mbox{sgn}}

\newfont{\bb}{msbm10 scaled 1100}

\begin{document}
\pagestyle{plain}

%\bstctlcite{BSTcontrol}
%\eqsec

\title{Tails of Random Matrix Diagonal Elements: \\The Case of the Wishart Inverse} %
%\title{SINR Distribution of Linear MIMO Receivers} %

\author{Aris L. Moustakas
\address{Physics Dept., National Capodistrian University of Athens\\ 157 84  Zografou, Athens, Greece} }
%\date{\today}

%\pubid{0000--0000/03\$17.00~\copyright~2003 IEEE}

\maketitle

\begin{abstract}
We analytically compute the large-deviation probability of a diagonal matrix element of two cases of random
matrices, namely $\beta=\left[\vec H^\dagger\vec H\right]^{-1}_{11}$ and $\gamma=\left[\vec I_N+\rho\vec H^\dagger\vec H\right]^{-1}_{11}$, where
$\vec H$ is a $M\times N$ complex Gaussian matrix with independent entries and $M\geq N$. These diagonal entries are related to
the ``signal to interference and noise ratio'' (SINR) in multi-antenna communications. They depend not only on the eigenvalues but also on the corresponding
eigenfunction weights, which we are able to evaluate
on average constrained on the value of the SINR. We also show that beyond a lower and upper critical value of $\beta$, $\gamma$,
the maximum and minimum eigenvalues, respectively, detach from the bulk. Responsible for this detachment is the fact that the
corresponding eigenvalue weight becomes macroscopic (i.e. $O(1)$), and hence exerts a strong repulsion to the eigenvalue.
\end{abstract}

%\PACS{PACS}

%%%%%%%%%%%%%%%%%%%%%%%%%%%
\section{Introduction}
\label{Introduction}
%%%%%%%%%%%%%%%%%%%%%%%%%%%

%\begin{itemize}
%\item This work:1: SNR special because is not a trace of matrix but just a matrix element. method can be generalized for {\em any}, also off-diagonal matrix element of a Wishart matrix.
%\item Interesting generalization is the joint distribution of two snr's. Away from the edges we can get an answer by integrating over $u$ and $v$ over all space. We should get estimates for edges etc.
%\end{itemize}

Random matrix theory has recently seen a flurry of applications in communications.
Here, the random matrix under study may be the matrix of random channel amplitudes between transmitting and receiving multi-antenna arrays,
 \cite{Foschini1998_BLAST1, Telatar1995_BLAST1}
or an array of pseudo-random code vectors used in a multi-user communications setting in order to scramble the signals from other
users \cite{Verdu_MUD_book, Verdu1999_MIMO1}.

One metric to characterize the performance of the communications is the mutual information, which gives the ultimate number of bits per channel use that
can be transmitted without error. The ergodic mean and the fluctuations of this quantity has been analyzed under a wide range of assumptions regarding the channel
statistics \cite{Telatar1995_BLAST1, Moustakas2000_BLAST1, Muller2002_RandomMatrixMIMO}. For slowly time-varying channels a better metric for the performance is the
so-called outage capacity which provides an achievable transmission rate given a probability that this rate will not be achieved from the underlying fading channel
\cite{Ozarow1994_OutageCapacity}. As a result, a number of works showed that for large antenna numbers the fading statistics become Gaussian
\cite{Moustakas2003_MIMO1, Hachem2006_GaussianCapacityKroneckerProduct}. More recently, the tails of the distribution were also calculated using the Coulomb Gas approach
\cite{Kazakopoulos2011_LivingAtTheEdge}.

To obtain the above full advantages from multiple antennas, it is necessary to have an optimal receiver structure, which however is quite complex to implement in real systems. Instead, low complexity albeit suboptimal linear receivers offer as a practical alternative.

Such receivers include the so-called MMSE (minimum mean square error) and the zero-forcing (ZF) receivers. The information throughput performance depends on the ability of the linear receiver structure to mitigate interference. One very useful method to quantify the performance is through the asymptotic analysis of the signal to interference and noise ratio (SINR) for the MMSE receiver in the limit of large antenna numbers using tools from random matrix theory.

As in the case of the mutual information, when the channel is slowly varying, it is important to evaluate the full probability
distribution of the SINR to obtain the probability of outage for a given target SINR value. This is important when the
number of antennas is not too large, in which case the fluctuations play an important role.

In a seminal work \cite{Tse2000_MMSEFluctuations} the authors proved the asymptotic normality of the SINR for the MMSE and ZF receivers when all transmitters have equal power.
 More recently, \cite{Liang2007_MMSEAsymptotics, Kammoun2009_CLT_MMSE_RMT} showed the normality of the MMSE SINR. Unfortunately and in contrast to the total mutual information, the Gaussian approximation for the SINR behaves badly unless the number of channels is quite large. As a result, inspired by the fact that the SINR for the equal power MIMO ZF receiver has a Gamma distribution\cite{Tse2000_MMSEFluctuations, Gore2002_MIMO_ZFReceiver}, several works were devoted in approximating the SINR statistics with other distributions, notably the Gamma and generalized Gamma probability densities \cite{Li2006_MIMO_MMSE_SINR_Distribution, Kammoun2009_BER_Outage_Approximations_MMSE_MIMO, Armada2009_BitLoadingMIMO, Li2011_BER_MIMO_MMSE}, by matching the first three moments. Nevertheless, this methodology, although perhaps providing good agreement under certain conditions, is ad-hoc and does not offer any intuition on the SINR statistics. Finally, it should be pointed out that the exact distribution of the SINR has been calculated recently in terms of ratios of determinants \cite{Kiessling2003_ExactMMSE_SINR}. Nevertheless, such an analysis
is quite tedious and does not provide any intuition about the result.

In this paper, we take a different approach. Instead of trying to prove Gaussian behavior close to the peak of the distribution of SINR, we apply the Coulomb Gas
methodology, which allows us to calculate the distribution of the SINR arbitrarily far from its most probable, ergodic value. The Coulomb gas model
was introduced originally by \cite{Dyson1962_DysonGas} and more recently has seen numerous applications \cite{Vivo2008_DistributionsConductanceShotNoise,
Dean2008_ExtremeValueStatisticsEigsGaussianRMT, Majumdar2006_LesHouches, Nadal2009_NonIntersectingBrowianInterfaces, Nadal2010_PhaseTransitionsBipartiteEntanglement,
Kazakopoulos2011_LivingAtTheEdge} treats each eigenvalue of a random matrix as a point charge in the presence of an external potential while repelling the others.
To apply this model, we rely on the fact that the SINR can be written in terms of a diagonal matrix of a random matrix and hence as a sum over the eigenvalues of the matrix.
Nevertheless,  since this sum depends not only on the eigenvalues but also the weights of the corresponding eigenfunctions on the matrix element,
we need to generalize the Coulomb gas approach to take into account the effects of the fluctuating weights. It should also be mentioned that single matrix element distributions
of related quantities have been evaluated elsewhere \cite{Savin2005_UniversalStatisticsGreenFunctionChaoticCavities},
but in a different context and without exemplifying the interaction between eigenvalues and eigenfunctions.

{\em Outline:}
In the next section we present the channel model and introduce the concepts of the SINR for the ZF and MMSE receiver.
In Section \ref{sec:technical} we present our analytical results, while in Section \ref{sec:numerical simulations} we demonstrate their validity numerically and we conclude in Section \ref{sec:conclusion}.

%%%%%%%%%%%%%%%%%%%%%%%%%%%%%%
\section{Problem Statement}
\label{sec:channel model}
%%%%%%%%%%%%%%%%%%%%%%%%%%%%%%

In this section we define the channel model. We consider a wireless communications system with an $N$ antenna transmitter array and an $M$ antenna receiver array. It is typically assumed that $M\geq N$ and the ratio is defined as $\alpha=M/N\geq 1$. The $M$-dimensional received signal vector $\vec y$ can be written as
\begin{equation}\label{eq:channel_def}
    \vec y = \vec H \vec x + \vec z
\end{equation}
where the vector $\vec x$ represents the transmitted signal with identically distributed elements and variance $\ex [\vec x\vec x^\dagger]=\rho \vec I_N$.  $\vec z$ is the  noise vector, with independent complex Gaussian elements $\sim {\cal CN}(0,1)$. The channel matrix $\vec H$ is assumed to have independent elements $\sim {\cal CN}(0,1/N)$.

The basic communications problem at the receiver is try to deduce $\vec x$ from $\vec y$, given the knowledge of the channel $\vec H$. Even though the optimal receiver structure leads to the maximum throughput per channel use there are several suboptimal receivers, which are popular because they are linear in their implementation. The most common ones are the so-called Minimum-Mean-Square-Error (MMSE) and the Zero-Forcing (ZF) receivers. In both cases the vector $\vec y$ is multiplied by a matrix $\vec P$ in an effort to mitigate the noise and interference.

\subsection{MMSE Receiver}
\label{sec:MMSE_receiver}

In this case the matrix $\vec P_{mmse}$ is
\begin{equation}\label{eq:mmse_projector}
    \vec P_{mmse}=\left[\vec I_M+\rho\vec H\vec H^\dagger\right]^{-1}\vec H^\dagger
\end{equation}
This matrix has the property of minimizing the average square error of the signal in the presence of the noise $\vec z$. The output signal can be then expressed as
\begin{equation}\label{eq:mmse_output}
    \vec {\hat x} = \vec P_{mmse} \vec y
\end{equation}
The resulting signal-to-interference-and-noise ratio (SINR) for each signal stream $x_i$ for $i=1,\ldots,N$ is given by \cite{Verdu_MUD_book}
\begin{eqnarray}\label{eq:SINR_mmse_def}
    \gamma_i &=& \frac{1}{\left[\left(\vec I_N +  \rho \vec H^\dagger \vec H\right)^{-1}\right]_{ii}} - 1
\end{eqnarray}
It will turn out to be convenient to parameterize this quantity by $z_i=\gamma_k/\rho$ and re-write the above equation as
\begin{eqnarray}\label{eq:SINR_mmse_def2}
    \frac{1}{1+\rho z_i} &=& \left[\left(\vec I_N +  \rho \vec H^\dagger \vec H\right)^{-1}\right]_{ii} \\
    &=& \sum_{j=1}^N \frac{\left|u_{ji}\right|^2}{1+\rho x_j}
\end{eqnarray}
In the second line we have expressed the $i$-th diagonal element of the matrix in the RHS of (\ref{eq:SINR_mmse_def2}) in terms of the eigenvalues $x_j$ and the matrix elements of the unitary matrix $\vec U$, which diagonalizes $\vec H^\dagger\vec H$. Since the elements of $\vec H$ are $\sim {\cal CN}(0,1/N)$, $\vec U$ is Haar-unitary matrix. As a result, the quantities $|u_{ji}|^2$ for fixed $i$ and $j=1,\ldots,N$ are uniformly distributed in $(0,1)$ with the constraint
\begin{eqnarray}\label{eq:u_ik_constraint}
\sum_{j=1}^N \left|u_{ji}\right|^2 = 1
\end{eqnarray}

\subsection{ZF Receiver}
\label{sec:ZF Receiver}

Similarly the SINR of the zero-forcing (ZF) receiver can be obtained. In this case the projector matrix $\vec P_{zf}$ is simply the pseudo-inverse of the matrix $\vec H$ ($\vec H^+$), which if course exists with probability one only when $M\geq N$. As a result, the  output vector is
\begin{equation}\label{eq:zf_output}
    \vec {\hat x} = \vec P_{zf} \vec y = \vec x + \vec H^+ \vec z
\end{equation}
We see that multiplication with $\vec H^+$ on $\vec y$ kills all self-interference of signals, since all terms involving signals $x_q$, with $q\neq i$ are forced to zero (hence the name ``zero-forcing''). Of course this comes at the cost of increasing the noise. The corresponding SINR $\beta_k$ can be written as
\begin{eqnarray}\label{eq:SINR_zf_def}
    \frac{\rho}{\beta_i} &=& \frac{1}{z_i} = \left[\left(\vec H^\dagger \vec H\right)^{-1}\right]_{ii} \\
    &=& \sum_{j=1}^N \frac{\left|u_{ji}\right|^2}{x_j}
\end{eqnarray}
with the quantities $u_{ji}$ and $x_j$ defined as above.

It is worth pointing out that the SINR of both MMSE and ZF cases above may be written as a sum over a function of eigenvalues weighted by the corresponding eigenvector weight.
\begin{eqnarray}\label{eq:SINR_gen_def}
    \sum_{j=1}^N \left|u_{ji}\right|^2 s(x_j)
\end{eqnarray}
Also, it is important to mention that in the limit of large $\rho$, $z_{mmse}$ coincides with $z_{zf}$, i.e. $z_{zf}=\lim_{\rho\rightarrow\infty} z_{mmse}$ and thus $\beta(\rho)=\rho \lim_{\rho'\rightarrow\infty} \gamma(\rho')/\rho'$. As a result, we will focus on the distribution of the MMSE SINR first,
from which we will be able to derive all results for the ZF SINR by taking the appropriate limit.

%%%%%%%%%%%%%%%%%%%%%%%%%%%%%%
\section{Technical Analysis}
\label{sec:technical}
%%%%%%%%%%%%%%%%%%%%%%%%%%%%%%

In this section we will go through the basic steps of the calculation of the probability distribution function (PDF)
of the normalized SINR $z_i$, omitting the index $i$ when necessary.
Keeping in mind the statistics of $u_{ji}$ and their constraint (\ref{eq:u_ik_constraint}) we write the PDF of $z$ as
\begin{eqnarray}\label{eq:PDF_def}
    \prob (z) = \frac{1}{N |s'(z)|}\ex_{\vec x, \vec t}\left[\delta\left(Ns(z)-\sum_{j=1}^N s(x_j) t_j\right)\right]
\end{eqnarray}
where the expectation is over the vector $\vec x$ of the $\vec H^\dagger \vec H$ eigenvalues and the random vector $\vec t$ with distribution
identical to the quantities $N|u_{ji}|^2$, for fixed $i$ and $j=1,\ldots,N$. We have used the compact notation $s(x)$ to indicate both ZF and MMSE cases above. For simplicity we will omit the dependence of $s(z)$ on $z$ as well as the overall proportionality factor unless explicitly mentioned.
We may explicitly integrate over $\vec t$ by first expressing the above $\delta$-function as well as the constraint
\begin{equation}\label{eq:t_constraint}
\sum_{j=1}^N t_j=N
\end{equation}
as Fourier integrals. As a result we obtain
\begin{eqnarray}\label{eq:PDF_def2}
    \prob (z) &\propto & \int dk \int d\lambda\ex_{\vec x}\left[e^{N(ks+\lambda)}\prod_{j=1}^N\int_0^N dt_j e^{-(\lambda+ks(x_j))t_j} \right] \\ \nonumber
    &\propto& \int dk \int d\lambda\ex_{\vec x}\left[e^{N(ks+\lambda)}\prod_{j=1}^N
    \left[\frac{1-e^{-N(ks(x_j)+\lambda)}}{(ks(x_j)+\lambda)}\right]\right]
\end{eqnarray}

Note that that although the integral over $k$ and $\lambda$ is along the imaginary line, the saddle point will lie on the real axis and hence we omit the imaginary $i$ for simplicity.
The expectation over $\vec x$ is performed with the eigenvalue distribution $P(\vec x)$ given by
\begin{eqnarray}\label{eq:P(x)_def}
    P(\vec x) &\propto & \Delta(\vec x)^2 \prod_{j=1}^N x_j^{M-N} e^{-Nx_j} \equiv e^{-N^2 F(\vec x)}
\end{eqnarray}
with $\Delta(\vec x)=\prod_{i>j}(x_i-x_j)$ the Vandermonde determinant and the second equation being the definition of $F(\vec x)$. When $N$ is large, the eigenvalues
of $\vec H^\dagger\vec H$ form a tight density which can be represented as a density $p(x)$ corresponding to the quantity
\begin{equation}\label{eq:p_x_def}
    p(x)=\frac{1}{N} \sum_j \delta(x-x_j)
\end{equation}
As a result we may rewrite (\ref{eq:PDF_def2}) as
\begin{eqnarray}\label{eq:PDF_def3}
    \prob (z) &\propto & \int dk d\lambda \int Dp e^{-N^2 F[p]} e^{-NE_0[p]}
\end{eqnarray}
where $\int Dp$ represents a path integral over non-negative, normalized $p(x)$, $F[p]$ is the energy functional associated with the probability distribution of
eigenvalues (\ref{eq:P(x)_def})
\cite{Vivo2007_LargeDeviationsWishart, Dean2008_ExtremeValueStatisticsEigsGaussianRMT, Kazakopoulos2011_LivingAtTheEdge} and $E_0[p]$ is the
functional obtained from the exponent in (\ref{eq:PDF_def})
\begin{eqnarray}\label{eq:F[p]}
    F[p] & = & \int_a^b dx p(x)\left(x-(\alpha-1)\ln(x) -\int_a^b dx' p(x')\ln|x-x'|\right) \\
    \label{eq:E[p]}
    E_0[p] &=&-(ks+\lambda)-\int_a^b dx p(x) \ln\left[\frac{1-e^{-N(ks(x)+\lambda)}}{k s(x) +\lambda}\right]
\end{eqnarray}
It is crucial to point out that the functional $F[p]$ in (\ref{eq:P(x)_def}) is multiplied by $N^2$ while $E_0[p]$ is only multiplied by $N$. Hence the fluctuations
of $F[p]$ will be far smaller than those of $E_0[p]$. As a result, to leading order in $N$ we may first find the optimal distribution that minimizes $F[p]$. This
distribution is the celebrated Marcenko-Pastur distribution given by \cite{Dean2008_ExtremeValueStatisticsEigsGaussianRMT, Vivo2007_LargeDeviationsWishart, Kazakopoulos2011_LivingAtTheEdge}
\begin{equation}\label{eq:MP_def}
p_0(x)=\frac{\sqrt{(x-a)(b-x)}}{2\pi x}
\end{equation}
where the limits of the support are
\begin{equation}\label{eq:ab_def}
    a,b = \left(\sqrt{\alpha}\pm1\right)^2
\end{equation}
Subsequently, using this $p_0(x)$ we may find the optimal values of $k$ and $\lambda$ that minimize $E_0[p_0]$. This two-tiered approach works, as mentioned before, because, for large $N$, the eigenvalue distribution has much smaller fluctuations compared to the fluctuations of the unitary matrix elements $|u_{ji}|^2$ and $k$, $\lambda$.

We next analyze the above equations in two separate regimes, depending on whether the quantity $\lambda+ks(x)$ is positive. When it is, the exponential factor
 inside the logarithm of (\ref{eq:E[p]}) is negligible, and we may therefore omit it. This corresponds to the situation when all weights of the eigenvalues are of similar size,
 i.e. $|u_{ji}|^2 = O(1/N)$, or equivalently $t(x_j)=O(1)$. The analysis of this region will be discussed next in Section \ref{sec:regiond>0}.
When $\lambda+ks(x)<0$, we need to take the exponential explicitly into account, which we will do in Section \ref{sec:regiond<0}. In this case, as we shall see, the
average weight of one eigenvalue becomes macroscopic, i.e. $t(x)=O(N)$.

\subsection{Region with $\lambda+ks(x)>0$}
\label{sec:regiond>0}

In this case, the exponential inside the logarithm  of (\ref{eq:E[p]}) is exponentially small in $N$ and therefore may be neglected.
As mentioned above, all typical values of $t_j$ are of order unity, i.e. $t_j=O(1)$, with their sum fixed to $N$  (\ref{eq:t_constraint}). In fact,
the  integral over $p(x)$ in (\ref{eq:E[p]}) represents the entropy of the random variables $\vec t$ for given $p(x)$ and the mean value of the $t_j=N|u_{ji}|^2$ is
equal to
\begin{equation}\label{eq:meanu_jk^2}
    t(x_j)=N\ex[|u_{ji}|^2]=\frac{1}{\lambda+ks(x_j)}
\end{equation}

The saddle point equations for $\lambda$ and  $k$ are obtained by differentiating $E_0[p_0]$ with respect to $\lambda$ and $k$, respectively, and setting the derivative to
zero
\begin{eqnarray}
    \label{eq:saddle_pt_eq_lambda}
    \int_a^b dx \frac{p_0(x)}{\lambda+ks(x)} &=& 1 \\
\label{eq:saddle_pt_eq_k}
    \int_a^b dx \frac{p_0(x)s(x)}{\lambda+ks(x)} & = & s(z)
\end{eqnarray}
By identifying $1/(ks(x_j)+\lambda)$ as the average value of $t_j$, we immediately see that the first equation is nothing else but the normalization condition (\ref{eq:t_constraint}). Similarly,
the second equation simply states that $\sum_j s(x_j) t_j = Ns$, i.e. imposes the $\delta$-function constraint in (\ref{eq:PDF_def}). We note that combining the two equations we get the identity
\begin{equation}\label{eq:ks+lambda=1}
    \lambda + ks(z)=1
\end{equation}
This, together with e.g. (\ref{eq:saddle_pt_eq_k}) will provide us with the optimal values of $k$, $\lambda$ to plug into (\ref{eq:E[p]})
and thus evaluate the leading term in the exponent of the probability distribution of $s$ (resp. $z$) by evaluating it at the saddle point.

%Before continuing into the details of the evaluation, it is worth discussing various limits of these equations. One important limit is $k=0$. In this case, (\ref{eq:ks+lambda=1})
% just states that $\lambda=1$, while (\ref{eq:saddle_pt_eq_k}) equates $s(z)$ to its ergodic average over the Marcenko-Pastur distribution, $\ex_p[s(x)]$.
%As we shall see, this corresponds to the peak of the Gaussian distribution of the SINR distribution. Of course, in the case of the ZF receiver, when $\alpha=1$ the
%integral will diverge, signalling the failure of the Gaussian approximation \cite{Tse2000_MMSEFluctuations, Kumar2009_LinearDMT}.
%First, since  $s(x)>0$,  the quantity $\lambda+ks(x)$ has to be positive in the support of $p(x)$. When this criterion fails, we may no longer neglect the exponential
%factor inside the logarithm of (\ref{eq:E[p]}) and need to include it in the calculation. It is easy to see that this criterion {\em needs} to breakdown for some values of $z$.
%Indeed,  since in both MMSE and ZF cases $s(x)$ is a decreasing function of $x$ we have the following inequality
%\begin{eqnarray}
%\label{eq:inequality_for_k}
%   s(b) =  s(b) \int_a^b dx \frac{p(x)}{\lambda+ks(x)} \leq s(z) \leq s(a) \int_a^b dx \frac{p(x)}{\lambda+ks(x)} = s(a)
%\end{eqnarray}
%As a result, for $z>b$ and $z<a$, (\ref{eq:saddle_pt_eq_k}) and (\ref{eq:saddle_pt_eq_c}) break down.

We start by making the following convenient change of variables from $k$  to $c$ through
\begin{equation}\label{eq:def_c}
    k=-\frac{\lambda}{s(c)}
\end{equation}
which for the MMSE case becomes $k= -\lambda(1+\rho c)$. Once this variable is determined, the other, e.g. $\lambda$ can be obtained from it through (\ref{eq:ks+lambda=1})
\begin{equation}\label{eq:soln_lambda}
    \lambda=\frac{s(c)}{s(c)-s(z)}
\end{equation}
Plugging (\ref{eq:def_c}) into (\ref{eq:saddle_pt_eq_k})
we get, after some re-arrangements,
\begin{eqnarray}
\label{eq:saddle_pt_eq_c}
    \int_a^b dx \frac{p_0(x)}{x-c} & = & \frac{1}{z-c}
\end{eqnarray}
where we have also used the fact that $s(z)=1/(1+\rho z)$. It is interesting to point out that this equation is independent of $\rho$ and holds also for the ZF receiver. Also it represents a balance of forces for a (yet fictitious) charge located at $c$: from one side
we have the repulsion of the Coulomb sea, while from the other there is another (fictitious) charge located at the position dictated by the normalized SINR $z$. Before proceeding to integrate the LHS we note that, in order to get a convergent answer,
$c$ has to take values outside the support of $p(x)$, i.e. $c\notin (a,b)$. We then have
\begin{eqnarray}
\label{eq:integral_c}
    z & = & \frac{\sgn(1+\alpha-c)\sqrt{(b-c)(a-c)}+c+\alpha-1}{2}
\end{eqnarray}
We see that the region of $z$ for which the above equation has solutions is $|z-\alpha|\leq \sqrt{\alpha}$. This corresponds to values of $c$ in the regions
$-\infty<c<a$ (for $\alpha-\sqrt{\alpha}<z<\alpha$) and $b<c<+\infty$ (for $\alpha<z<\alpha+\sqrt{\alpha}$).
Solving the above equation for $c$ gives us
\begin{equation}\label{eq:soln_c}
    c(z)=z\left(1+\frac{1}{z-\alpha}\right)
\end{equation}
The values of $c(z)$ for which the solutions above break down are $c=a$ ($z(a)=\alpha-\sqrt{\alpha}$) and $c=b$ ($z(b)=\alpha+\sqrt{\alpha}$).

We may now calculate the exponent of the PDF for $|z-\alpha|\leq \sqrt{\alpha}$. To do so, we simply need to plug in the above values
of $k$ and $\lambda$ (obtained directly from $c$) into (\ref{eq:E[p]}) and calculate the corresponding integrals. As discussed before \cite{Dean2008_ExtremeValueStatisticsEigsGaussianRMT, Vivo2007_LargeDeviationsWishart, Kazakopoulos2011_LivingAtTheEdge}
the value of $F[p_0]$ does not depend on $z$ and is therefore a constant (we have also omitted the dependence of $E_0$ on $p_0$).
\begin{eqnarray}
\label{eq:E[p]_soln_inner}
    E_0(z) &=& -\ln\left[\frac{z-c}{\rho^{-1}+z}\right]+\frac{c+\rho^{-1}}{2} \\ \nonumber
    &+& \frac{1}{2}\left(\sgn(1+\alpha-c)\sqrt{(b-c)(a-c)}-\sqrt{(\rho^{-1}+a)(\rho^{-1}+b)}\right)  \\ \nonumber
    &+&(\alpha+1)\ln\left[\frac{\sqrt{|b-c|}+\sqrt{|a-c|}}{\sqrt{\rho^{-1}+b}+\sqrt{\rho^{-1}+a}}\right] \\ \nonumber
    &-&(\alpha-1)\ln\left[\frac{\sqrt{a|b-c|}+\sqrt{b|a-c|}}{\sqrt{a(\rho^{-1}+b)}+\sqrt{b(\rho^{-1}+a)}}\right]
\end{eqnarray}
Plugging in the dependence of $c(z)$ we obtain the following simplified formula.
\begin{eqnarray}
\label{eq:E[p]_soln_inner_simple}
    E_0(z) &=& z-\alpha\ln z + \ln(\rho^{-1}+z)+\frac{\rho^{-1}+1-\alpha}{2} \\ \nonumber
    &-&\frac{\sqrt{(\rho^{-1}+a)(\rho^{-1}+b)}}{2} + \ln 4 +\frac{(\alpha+1)\ln\alpha}{2} \\ \nonumber
    &-&(\alpha+1)\ln\left[\sqrt{\rho^{-1}+a} +\sqrt{\rho^{-1}+b}\right] \\ \nonumber
    &+& (\alpha-1)\ln\left[\sqrt{b(\rho^{-1}+a)} +\sqrt{a(\rho^{-1}+b)}\right]
\end{eqnarray}
This exponent has two interesting properties. First, it is a maximum at the ergodic value of $z$, which corresponds to the ergodic average of the SINR. This corresponds to $k=0$, and,
following (\ref{eq:ks+lambda=1}) also $\lambda=1$. As a result, (\ref{eq:saddle_pt_eq_k}) equates $s(z)$ to its ergodic average over the Marcenko-Pastur distribution,
$\ex_{p_0}[s(x)]$. As we shall see, this corresponds to the peak of the Gaussian distribution of the SINR distribution.
The MMSE SINR is the given by
\begin{eqnarray}
\label{eq:MMSE_SINR_erg}
    \gamma_{erg} &=& \rho z_{erg,mmse} = \frac{\sqrt{(1-(\alpha-1)\rho)^2+4\alpha\rho}+\rho(\alpha-1)-1}{2}
\end{eqnarray}
This can be seen by directly maximizing (\ref{eq:E[p]_soln_inner_simple}) over $z$. Indeed, expanding (\ref{eq:E[p]_soln_inner_simple}) close to $z_{erg}$ we find
\begin{eqnarray}
\label{eq:MMSE_SINR_gauss}
    E_0(z) &\approx& \frac{\left(z-z_{erg}\right)^2}{2 v_{erg}}
\end{eqnarray}
where $v_{erg}$ is the variance of the MMSE SINR given by \cite{Kumar2009_LinearDMT}
\begin{eqnarray}
\label{eq:MMSE_SINR_var_erg}
    v_{erg} &=& \frac{(\alpha-1)\sqrt{(1-(\alpha-1)\rho)^2+4\alpha\rho}+\rho(\alpha-1)+\alpha+1}{2\sqrt{(1-(\alpha-1)\rho)^2+4\alpha\rho}}
\end{eqnarray}
The corresponding values of $\beta_{erg}$ and $v_{erg}$ for the case of the ZF receiver can be obtained by taking the limit $\rho\rightarrow\infty$ but keeping $z$ fixed,
and then multiplying with $\rho$ to get the SINR $\beta_{erg}$
\begin{eqnarray}
\label{eq:ZF_SINR_erg}
    \beta_{erg} &=& \rho z_{erg,zf} = \rho(\alpha-1) \\
    v_{erg,zf} &=& \alpha-1
\end{eqnarray}
We see that when $\alpha=1$ the Gaussian approximation breaks down \cite{Tse2000_MMSEFluctuations, Kumar2009_LinearDMT}.

Keeping only the dependence  on $z$ in (\ref{eq:E[p]_soln_inner_simple}), we see that, to leading exponential order,
\begin{eqnarray}\label{eq:PDF_simple}
    \prob_{mmse} (z) &\propto & \frac{z^M}{(\rho^{-1}+ z)^N}e^{-Nz} \\ \nonumber
    \prob_{mmse}(\gamma) &\propto& \frac{\gamma^M}{(1+\gamma)^N}e^{-N\gamma/\rho}
\end{eqnarray}
This formula is remarkable for two reasons. First, it is surprising that this simple formula peaks at the ergodic value of $z_{erg}$ in (\ref{eq:MMSE_SINR_erg}). Second, it
settles a year-old conjecture, that the distribution of MMSE SINR should be (approximately) a Gamma distribution. Several papers in the literature \cite{Li2006_MIMO_MMSE_SINR_Distribution,
 Armada2009_BitLoadingMIMO, Kammoun2009_BER_Outage_Approximations_MMSE_MIMO} tried to fit the distribution to the Gamma distribution by fitting their moments. We see that this asymptotic form
illustrates that although simple in form, it is not a Gamma distribution.

By letting $\rho\rightarrow\infty$ we can recover the distribution of $z_{zf}$
 \begin{eqnarray}\label{eq:PDF_ZF_simple}
    \prob_{zf} (z) &\propto & z^{M-N}e^{-Nz} \\ \nonumber
    \prob_{zf}(\beta) &\propto& \beta^{M-N}e^{-N\beta/\rho}
\end{eqnarray}
It turns out that this result is in fact exact \cite{Tse2000_MMSEFluctuations, Gore2002_MIMO_ZFReceiver}.

Finally, we can also evaluate the average weight of each eigenvalue $x\in[a,b]$ constrained on the value of $z$. Using (\ref{eq:meanu_jk^2}), (\ref{eq:soln_lambda})
and (\ref{eq:soln_c}) we obtain
\begin{equation}\label{eq:meanu_jk^2_explicit}
    \ex[t(x|z)] = \frac{1}{\lambda+ks(x)}=\frac{s(z)-s(c)}{s(x)-s(c)}
\end{equation}
for $|z-\alpha|<\sqrt{\alpha}$, which is valid for both ZF and MMSE. This result, can also be obtained by noting that the distribution of $t(x|z)$ is exponential
$\sim \exp[-(\lambda+ks(x))t]$, as seen in (\ref{eq:PDF_def2}).

\subsection{Region with $\lambda+ks(x)\leq 0$}
\label{sec:regiond<0}

Before moving on, it is worth pointing out that the above behavior is bound to break down at {\em some} value of $z$.
Indeed, assuming $\lambda+ks(x)>0$ and using (\ref{eq:saddle_pt_eq_lambda}), (\ref{eq:saddle_pt_eq_k}) and the fact that $s(x)$ is a decreasing function of $x$,
we get the following inequality
\begin{eqnarray}
\label{eq:inequality_for_k}
   s(z) &=&   \int_a^b dx \frac{p(x)s(x)}{\lambda+ks(x)} \leq s(a) \\ \nonumber
   s(z) &=&   \int_a^b dx \frac{p(x)s(x)}{\lambda+ks(x)} \geq s(b)
\end{eqnarray}
Therefore, for $z<a$ and $z>b$, the assumption $\lambda+ks(x)>0$ and the resulting equations (\ref{eq:saddle_pt_eq_lambda}) and (\ref{eq:saddle_pt_eq_k}) have to break down.

To see how, we need to analyze the situation outside the region $|z-\alpha|<\sqrt{\alpha}$.
We thus need to consider the situation when for some eigenvalue(s) the exponent in
(\ref{eq:E[p]}) becomes positive, i.e. when $\lambda+ks(x)<0$. For the section of the support of $p(x)$ where this occurs, the exponent in (\ref{eq:E[p]}) will be positive, so we will need to include an additional term in $E_0[p]$, namely
\begin{eqnarray}\label{eq:E[p]_ks+l<0}
    NE_0[p] &\approx &-(ks(z)+\lambda)N+N\int_a^b dx p(x) \ln\left|k s(x) +\lambda\right| \\ \nonumber
    &+&N^2\int_{\cal R} dx  p(x) (ks(x)+\lambda)
\end{eqnarray}
where ${\cal R}$ is the region of the support of $p(x)$ (possibly including only a finite number of
eigenvalues) with $\lambda+ks(x)<0$. This extra term is an additional potential of strength $N^2 k s(x)$
exerting a force on the charge density $p(x)$. Since it is $O(N^2)$ we can no longer assume it is small and we have to take it into account explicitly together with $N^2F[p]$ in the determination of the optimal $p(x)$.
First, we need to estimate whether the number of eigenvalues affected is finite or scales with $N$. To answer this we start by observing that for these eigenvalues
the corresponding typical value of $t_j$ becomes of $O(N)$. Due tot the constraint $\sum_j t_j=N$
(\ref{eq:t_constraint}), there can be at
most a finite number of such eigenvalues with corresponding $t_j=O(N)$. We will initially assume that it is only one such eigenvalue
and later on show that this is consistent. As a result of these considerations
we need to treat this eigenvalue separately from the others. Therefore, we separate the continuous part of the eigenvalue density and express is as
\begin{equation}\label{eq:cont_density_N-1}
    q(x) = \frac{1}{N-1}\sum_{j=1}^{N-1} \delta(x-x_j)
\end{equation}
and denote the position of the $N$ eigenvalue (which may be the largest or smallest depending on whether we are analyzing the case $z>\alpha+\sqrt{\alpha}$
or $z<\alpha-\sqrt{\alpha}$, respectively) by $y$. The exponent in  (\ref{eq:PDF_def3}) can be expressed as
\begin{eqnarray}\label{eq:PDF_def4_N-1}
    -(N-1)^2 F[q]-NE_+[q]
\end{eqnarray}
where $F[q]$ is the same energy functional as in (\ref{eq:PDF_def3}). As a result, the optimal $q(x)$ is still the Marcenko-Pastur distribution (\ref{eq:MP_def}).  Thus $E_+[p_0]$ is given by
\begin{eqnarray}\label{eq:E+[p]}
    E_+[p,y] &=&y-(\alpha-1)\ln y-2\int_a^b dx p_0(x)\ln|x-y|-(ks(z)+\lambda) \\ \nonumber
    &+&\int_a^b dx p_0(x) \ln (ks(x)+\lambda) - \int_a^b dx p_0(x) \ln\left[1-e^{-N(ks(x)+\lambda)}\right] \\ \nonumber
    &+&\frac{1}{N}\left(\ln|\lambda+ks(y)|-\ln \left[e^{-N(ks(y)+\lambda)}-1\right]\right)
\end{eqnarray}
This need of explicitly splitting one eigenvalue from the bulk and treating it in a special way has appeared also in
 the context of bipartite entanglement \cite{Nadal2010_PhaseTransitionsBipartiteEntanglement,
Nadal2011_QuantumEntanglementRandomPureState}.
Since only $ks(x)+\lambda\geq 0$ for $x\neq y$ the last term in the second line above will only contribute subleading terms and therefore may be neglected.
We now need to find the saddle point jointly for $y$, $\lambda$ and $k$.
\begin{eqnarray}
    \label{eq:saddle_pt_eq_lambda2}
    1 &=& \int_a^b dx \frac{p_0(x)}{\lambda+ks(x)} + \frac{1}{N(\lambda+ks(y))} + \frac{e^{-N(\lambda+ks(y))}}{e^{-N(\lambda+ks(y))}-1}\\
\label{eq:saddle_pt_eq_k2}
    s(z) &=& \int_a^b dx \frac{p_0(x)s(x)}{\lambda+ks(x)} + \frac{s(y)}{N(\lambda+ks(y))} + \frac{s(y)e^{-N(\lambda+ks(y))}}{e^{-N(\lambda+ks(y))}-1}\\
\label{eq:saddle_pt_eq_y2}
    1 & = & 2\int_a^b dx \frac{p_0(x)}{y-x}+\frac{\alpha-1}{y} \\ \nonumber
    &+& \frac{k\rho s(y)^2}{N(\lambda+ks(y))} + \frac{k\rho s(y)^2e^{-N(\lambda+ks(y))}}{e^{-N(\lambda+ks(y))}-1}
\end{eqnarray}
From the first equation we conclude that the values of $\lambda+ks(y)=O(1/N)$. Otherwise, if $\lambda+ks(y)=O(1)$, the integral in the
 right-hand-side of (\ref{eq:saddle_pt_eq_lambda2}) would have to vanish, which is inconsistent with the fact that $\lambda+ks(x)>0$. Thus, setting $\lambda+ks(y)=-w/N$
 for $w$ still unknown, we find that to leading order,
\begin{equation}\label{eq:k_soln2}
    k=\frac{1}{s(z)-s(y)}
\end{equation}
and $w$ is a solution of the equation
\begin{equation}\label{eq:a_soln2}
    \int_a^b dx \frac{p_0(x)(s(z)-s(y))}{s(x)-s(y)}=\frac{1}{w}-\frac{1}{e^w-1}
\end{equation}
Putting this together, we finally get the equation for $y$:
\begin{eqnarray} \label{eq:saddle_pt_eq_y3}
    \int_a^b \frac{p(x)dx}{y-x} = 1-\frac{\alpha-1}{y} - \frac{1}{y-z}
\end{eqnarray}
This is last equation is both surprising and intuitive. It is firstly surprising that the value of $y$ does not depend on the specific form of the SINR function $s(z)$
but only on $z$ itself, the normalized SINR. Second, it tells us that the position of this eigenvalue is determined by external forces exerted on the other eigenvalues and,
in addition, it feels the repulsion of a unit charge located in the position $z$. Also it is interesting to note that it feels only {\em half} the repulsion from the
Marcenko-Pastur continuous eigenvalue density. The other half has been {\em screened} away due to
the interaction of this eigenvalue with the matrix elements of the diagonalizing unitary matrix (through $\lambda$ and $k$).
It should be pointed out that (\ref{eq:saddle_pt_eq_y3}) is valid for both cases $z<\alpha-\sqrt{\alpha}$ and $z>\alpha+\sqrt{\alpha}$, with $y\leq a$ and $y>b$, respectively.

In the above analysis we have assumed there is only one eigenvalue that detaches from the bulk. Let us assume there were $r>1$ such eigenvalues.
In that case, they would all have to stick together satisfying $\lambda+ks(y_j)=-w_j/N$. Otherwise, if say only one $y_1$ satisfied this relation,
all others with $y_j-y_1=O(1)$, for $j=2,\ldots,r$,  would necessarily have $\lambda+ks(y_j)>0$, which would not be sufficient to provide the ``kick'' to get out of the
bulk. However, on the other hand, if these $r$ eigenvalues are within $O(1/N)$ from each other, their repulsion $1/(y_i-y_j)$ will be large ($O(N)$) and
hence would dominate (\ref{eq:saddle_pt_eq_y2}). As a result, only one eigenvalue can be detached from the bulk.

We may now integrate the LHS of (\ref{eq:saddle_pt_eq_y3}) to get
\begin{eqnarray} \label{eq:saddle_pt_eq_y4}
    \frac{\sqrt{(a-y)(b-y)}+y-\alpha+1}{2y} = 1-\frac{\alpha-1}{y} - \frac{1}{y-z}
\end{eqnarray}
Solving for $y$ gives
\begin{eqnarray} \label{eq:y_soln}
    y &=& z\left(1+\frac{1}{z-\alpha}\right) \\ \nonumber
    |z-\alpha|&>&\sqrt{\alpha}
\end{eqnarray}
which is identical with (\ref{eq:soln_c}), although obtained through a completely different method, and with $z$ here taking different values.
One way to jointly interpret $c$ and $y$ is that they correspond to the location of a ``state'', which when located outside the continuum of nearby states
forms a bound state ($y$), while when it enters the continuum, it becomes a ``resonance''.

We may now plug in the above results into (\ref{eq:E+[p]}) to obtain the exponent of the PDF in this region of $z$. The final result is
\begin{eqnarray}
\label{eq:E+[p]_soln_outer}
    E_+(z) &=& -\ln\left[\frac{|z-y|}{\rho^{-1}+z}\right]+\frac{y+\rho^{-1}}{2} -(\alpha-1)\ln y\\ \nonumber
    &+& \frac{1}{2}\left(\sgn(y-\alpha)\sqrt{(y-b)(y-a)}-\sqrt{(\rho^{-1}+a)(\rho^{-1}+b)}\right)  \\ \nonumber
    &-&(\alpha+1)\left(\ln\left[\sqrt{|y-b|}+\sqrt{|y-a|}\right]+\ln\left[\sqrt{\rho^{-1}+b}+\sqrt{\rho^{-1}+a}\right]\right) \\ \nonumber
    &+&(\alpha-1)\left(\ln\left[\sqrt{a|b-c|}+\sqrt{b|a-c|}\right]\right. \\ \nonumber
    &&+\left.\ln\left[\sqrt{a(\rho^{-1}+b)}+\sqrt{b(\rho^{-1}+a)}\right]\right)
\end{eqnarray}
Inserting (\ref{eq:y_soln}) into (\ref{eq:E+[p]_soln_outer}) we find that, up to a constant, $E_+(z)$ is identical to $E_0(z)$ in (\ref{eq:E[p]_soln_inner_simple}), thus extending the validity of the
latter for all values of $z>0$.

Finally, we evaluate the average eigenvalue weights constrained on the value of SINR (or equivalently to $z$), which are the analogues of
(\ref{eq:meanu_jk^2_explicit}). Using (\ref{eq:k_soln2}) we find the identical expression as (\ref{eq:meanu_jk^2_explicit})
\begin{equation}\label{eq:meanu_jk^2_explicit_l+ks>0}
    \ex[t(x|z)] = \frac{1}{\lambda+ks(x)} = \frac{s(z)-s(y)}{s(x)-s(y)}
\end{equation}
In addition, we may calculate the mean weight of the detached eigenvalue $y$. As with the other weights, its distribution is exponential $\sim \exp[-t(\lambda+ks(y))]$,
(\ref{eq:PDF_def2}). We thus find that in this case
\begin{eqnarray}\label{eq:meanu_extremum}
    \ex[|u_{Nk}|^2] &=& \frac{\ex[t_y(z)]}{N} = 1 -\frac{1}{w} + \frac{1}{e^w-1} \\ \nonumber
    &=& 1-\frac{z((1+\rho\alpha)(z-\alpha)+\rho\alpha)}{(1+\rho z)(z-\alpha)^2(z-\alpha+1)}
\end{eqnarray}
 for $|z-\alpha|>\sqrt{\alpha}$, where $w=-N(\lambda+ks(y))$ appears in (\ref{eq:a_soln2}). We see that in the limit $|z-\alpha|=\sqrt{\alpha}$, $\ex[t_y(z)]=0$ (i.e. $=O(1/N)$). We plot

\begin{figure*}%[htbp]
\begin{center}
\subfigure[Weight for ZF]
{\includegraphics[width=0.49 \textwidth]{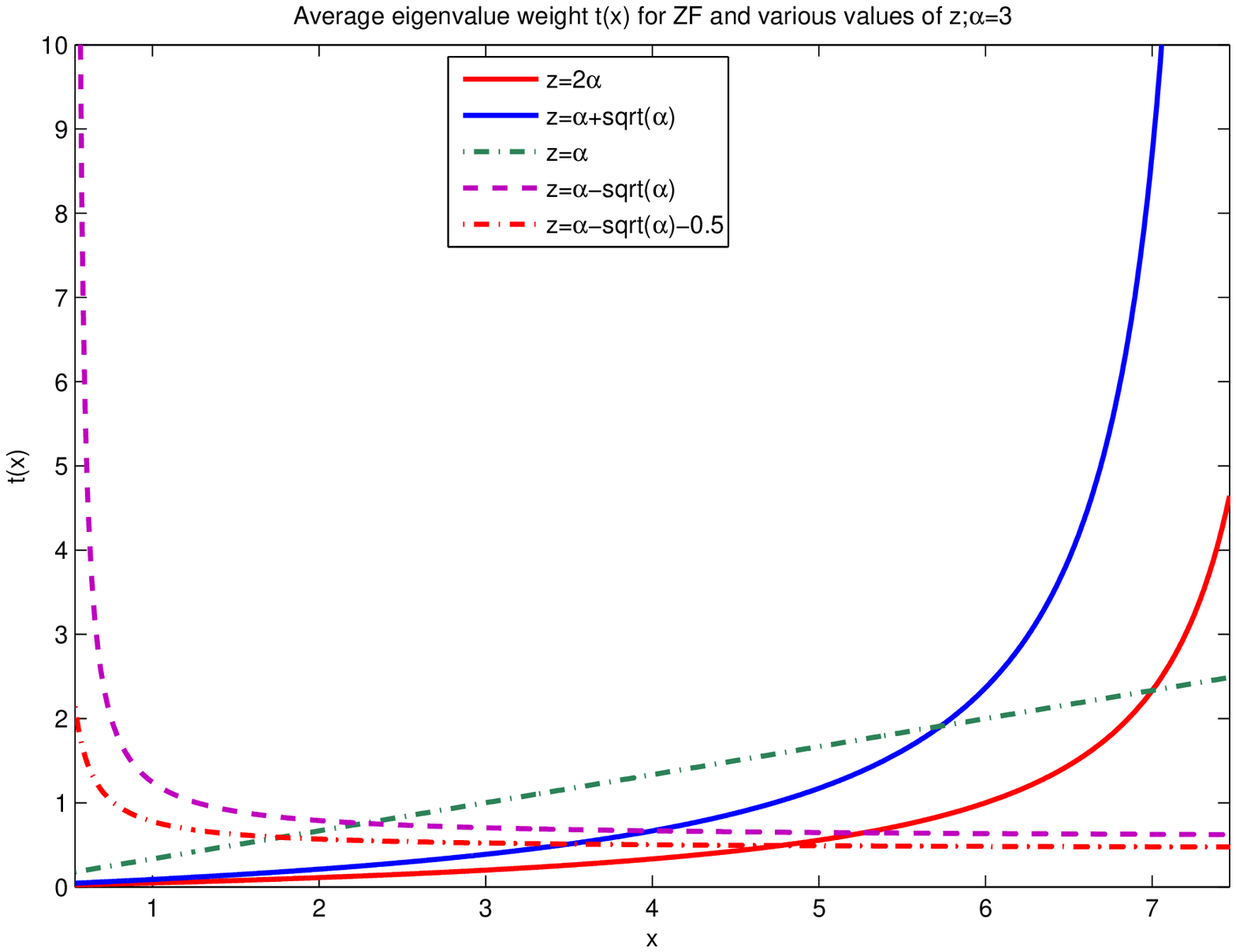}}
\subfigure[Weight of $x_{min}$]
{\includegraphics[width=0.49 \textwidth]{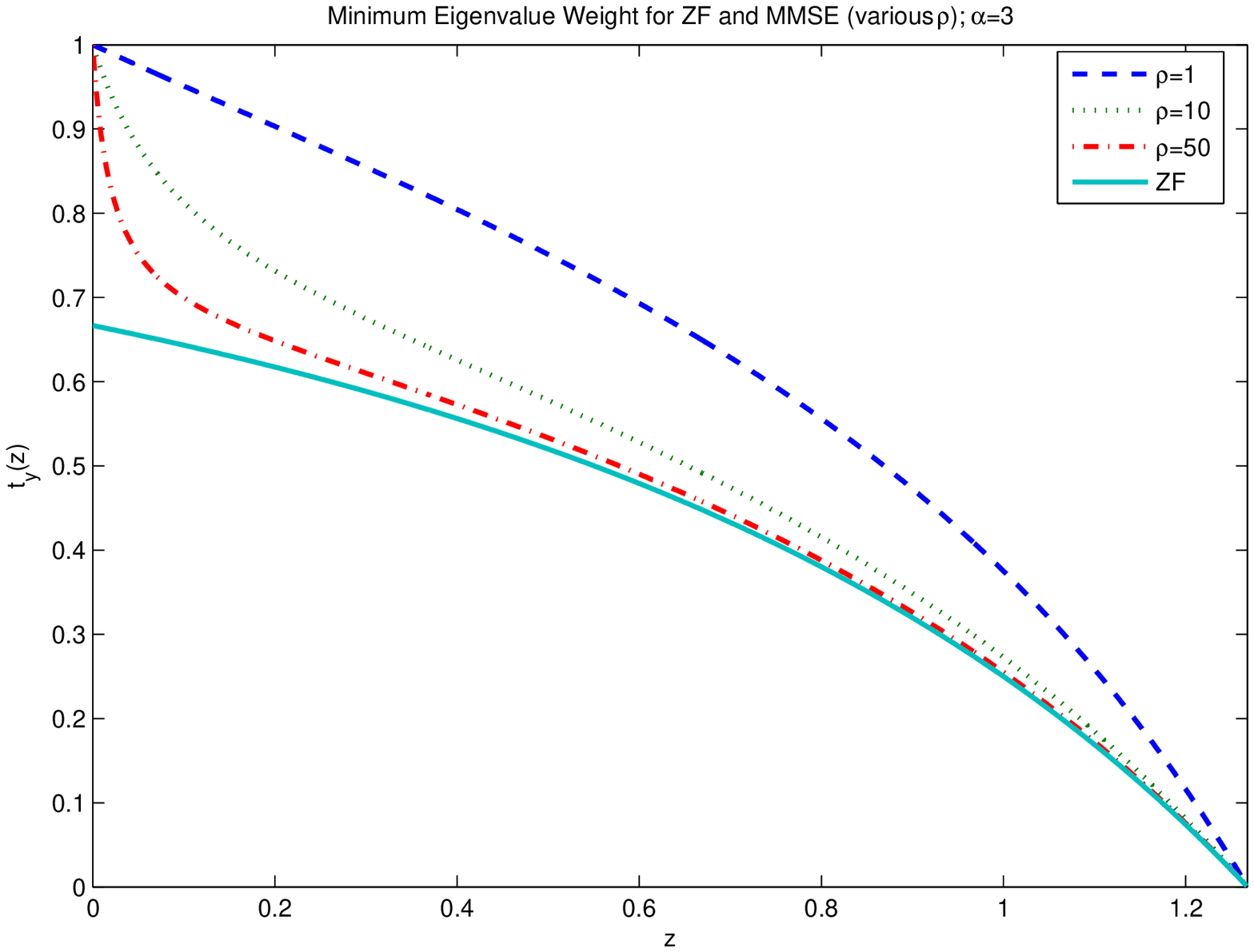}}
\caption{(a) Average weight of the eigenfunction $j$ on the $i$ element, $\ex [N|u_{ji}|^2]$ for the ZF case, as a
function of the corresponding eigenvalue $x_j$, constrained on different values of $z$.
The five curves include to the lower critical ($z=\alpha-\sqrt{\alpha}$), upper critical ($z=\alpha+\sqrt{\alpha}$) and
 an intermediate ($z=\alpha$) value of $z$. The remaining two curves have $z$ below the lower critical value $z=\alpha-\sqrt{\alpha}-0.5$ and above the upper critical value
 $z=2\alpha$.  In the first two we clearly see the divergence at the lower and higher edges of the
 spectrum, corresponding to the fact that beyond these values the weight of the edges becomes macroscopic $O(N)$. (b) Average weight of the minimum
 eigenvalue for $0<z<\alpha-\sqrt{\alpha}$. Note that now we plot the {\em macroscopic} occupation of the eigenvalue, i.e. $\ex [|u_{min,k}|^2]$. We plot the
 case of MMSE for various $\rho$ as well as the case of ZF.}
 \end{center}
\label{fig:t_x_different_z}
\end{figure*}

\begin{figure*}%[htbp]
\begin{center}
\subfigure[$\alpha=1$]
{\includegraphics[width=0.49 \textwidth]{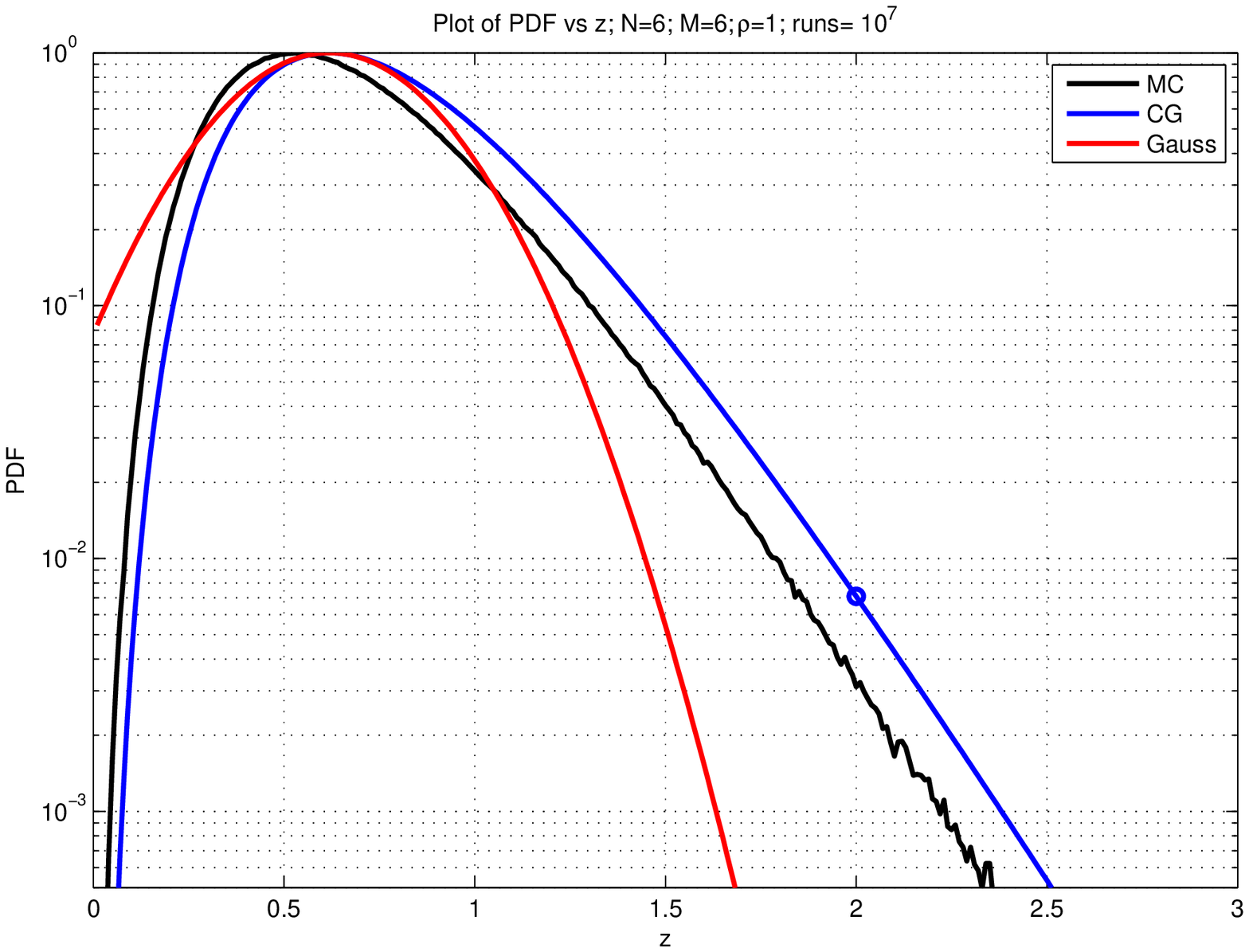}}
\subfigure[$\alpha=2$]
{\includegraphics[width=0.49 \textwidth]{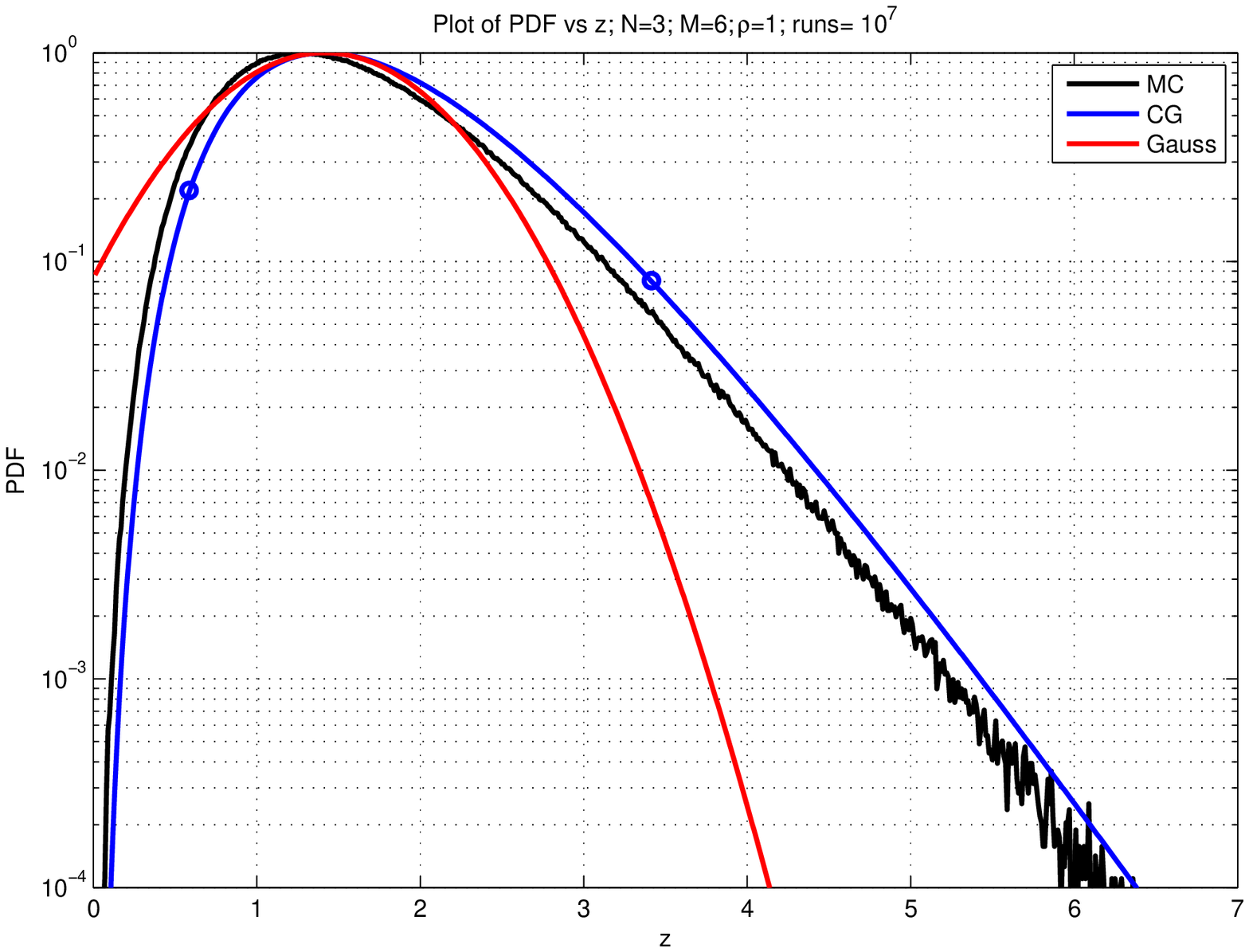}}
\caption{PDF of MMSE SINR for $M=6$, $\rho=1$. The agreement of the Coulomb Gas (CG) curve with Monte-Carlo (MC) simulations is good, even for such small matrices,
especially compared to the Gaussian approximation.
Denoted with circles are the values of $z$ at which the ``inner'' and ``outer'' solutions match and we see no discontinuity in the numerics. }
\end{center}
\label{fig:PDF_N36M6_rho1}
\end{figure*}

\begin{figure*}%[htbp]
\begin{center}
\subfigure[$\alpha=1$]
{\includegraphics[width=0.49 \textwidth]{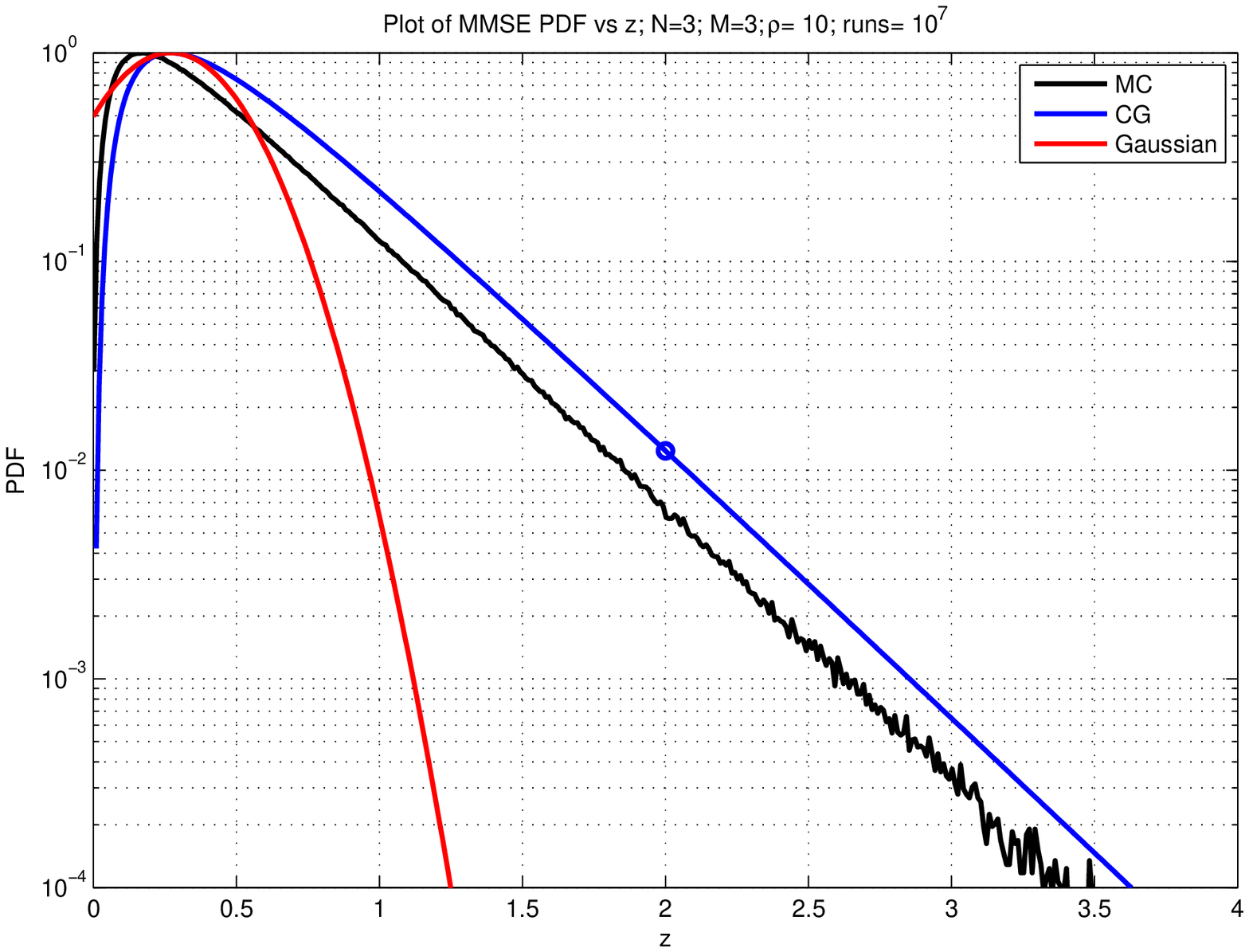}}
\subfigure[$\alpha=2$]
{\includegraphics[width=0.49 \textwidth]{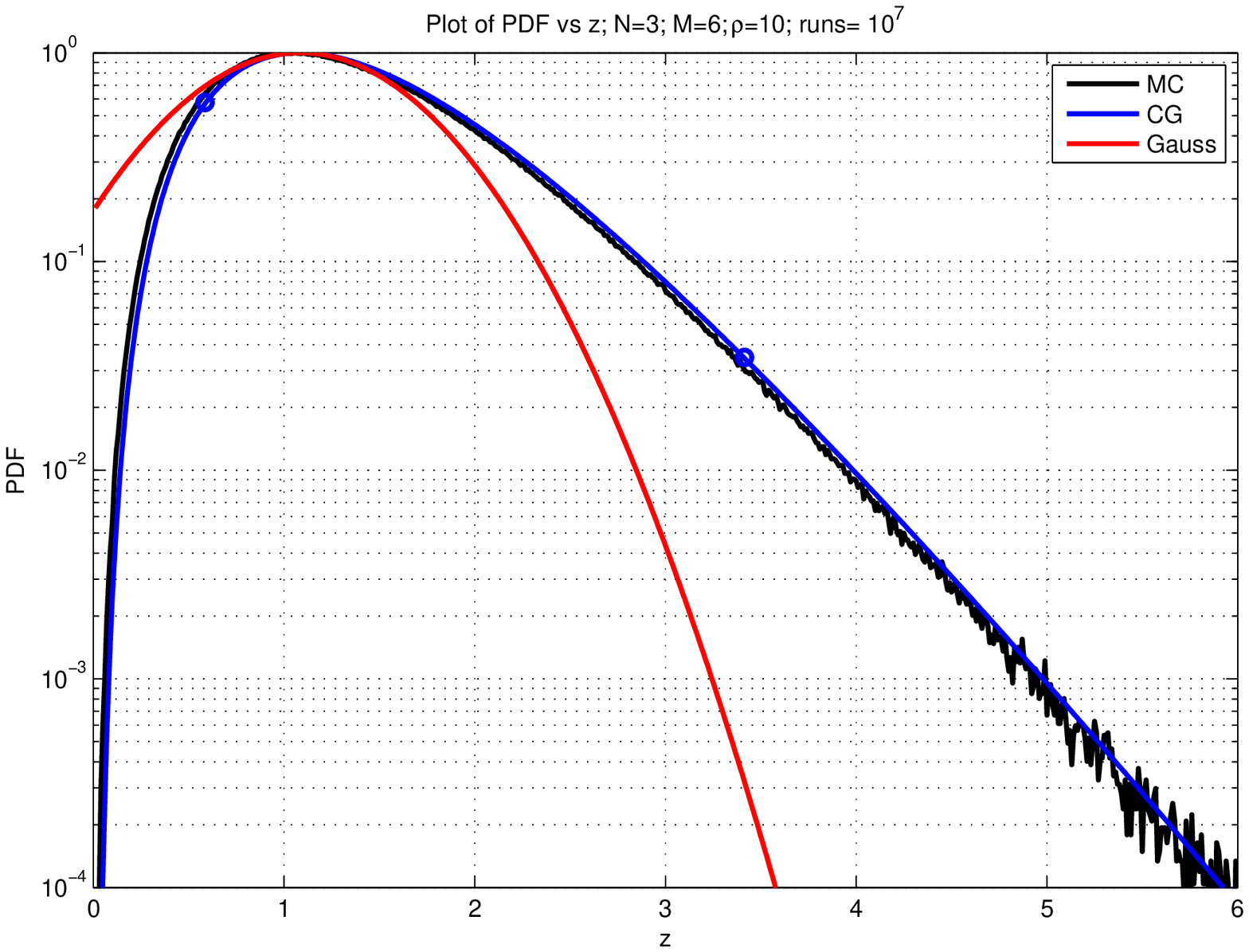}}
\caption{Same as in the previous figures but for $\rho=10$. Here the agreement is much better. }
\end{center}
\label{fig:PDF_N3M36_rho10}
\end{figure*}

\section{Numerical Simulations}
\label{sec:numerical simulations}

To test the applicability of this approach, we have performed Monte-Carlo simulations and have compared our large deviations Coulomb Gas (CG) approach with Monte Carlo
 (MC) simulations and the Gaussian approximation. In Fig. 2 %\ref{fig:PDF_N36M6_rho1}
 we plot the normalized probability density of the SINR for small $\rho=1$, while in
 Fig. 3%\ref{fig:PDF_N3M36_rho10}
we plot it for larger $\rho=10$. In both cases we see good agreement.

\section{Conclusion}
\label{sec:conclusion}

In this paper we have used a large deviation approach to calculate the probability density of the ``signal to interference and noise ratio'' (SINR) for multi-antenna arrays for two popular receiving algorithms,
namely the MMSE and the ZF algorithms. The approach is formally valid for large $N$ antenna numbers, but is not restricted to the behavior close to
the peak of the distribution, which has been shown to be asymptotically Gaussian when the number of antennas is very large. Instead we calculate the probability of the SINR being arbitrarily away from its ergodic
peak. Surprisingly, the leading term of the exponent of the distribution is very simple, and the distribution is neither Gamma, nor Beta and certainly not Gaussian. In the
ZF case, we recover the known chi-square result. We also test the MMSE results numerically and find good agreement even for relatively small antenna arrays. From a technical point of view, since the SINR of the
two algorithms are related to the diagonal matrix elements of the matrices  $\left[\vec I_N+\rho\vec H^\dagger \vec H\right]^{-1}$ and  $\left(\vec H^\dagger \vec H\right)^{-1}$,
the task is to find the distribution of a single diagonal matrix element.  The methodology we
applied is based on the so-called Coulomb Gas model, in which each eigenvalue can be seen as a point charge interacting with an external potential and repelling each other.
In this particular case however, the eigenvalues interact not only with each other but also with the weights of their corresponding eigenfunctions in the particular matrix element.
As a byproduct of our analysis we are able to calculate the average weight of each eigenvalue in the particular matrix element, constrained on the value of the SINR or the matrix element.
The interaction between the eigenfunction weights and the corresponding eigenvalues can be quite strong and as a result, below and above critical values of $z$ the lowest and
largest eigenvalues detach from the bulk. Nevertheless, it seems that there is no discontinuity involved in this detachment, at least to leading order. In hindsight, this is not
surprising. A given diagonal matrix element depends on a number of $O(N)$ random variables of the matrix, which has $O(N^2)$ random variables. In our approach
we have expressed this diagonal matrix in terms of the eigenvalues, which depend on the whole matrix. Somehow, we expect that the interaction with the eigenvalue weights will
``wash'' out this dependence from the full matrix.

\section*{Acknowledgments}
The author is grateful for the hospitality of the Jagellonian University in Krakow, Poland where the ideas of this work were formulated. He would also like to acknowledge insightful discussions with S. N. Majumdar that took place during the 23rd Marian Smoluchowski Symposium on
``Random Matrices, Statistical Physics and Information Theory'', Krak\'{o}w,
Poland, 26-30 Sept. 2010. Also, it is a pleasure to acknowledge useful discussions with G. Caire and P. Kazakopoulos  during the initial stages of the work.

%\appendices

%\bibliographystyle{IEEEtran}
%\bibliography{IEEEabrv,../../bibliography/wireless}


%\pagebreak

\end{document}